# The Case for Optical Interferometric Polarimetry

Nicholas M. Elias II[1,2], Carol E. Jones[3], Henrique R. Schmitt[4],
Anders M. Jorgensen[5], Michael J. Ireland[6], Karine Perraut[7]

[1]Zentrum für Astronomie der Universität Heidelberg, Landessternwarte; Königstuhl 12; 69117 Heidelberg; Germany
[2]Max-Planck-Institut für Astronomie; Königstuhl 17; 69117 Heidelberg; Germany
[3]University of Western Ontario; Department of Physics and Astronomy; London, ON N6A 3K7; Canada
[4]Naval Research Laboratory; Remote Sensing Division Code 7210; 4555 Overlook Avenue, SW; Washington, DC 20375; USA
[5]New Mexico Institute of Mining and Technology; Department of Electrical Engineering; 801 Leroy Place; Socorro, NM 87801; USA
[6]Sydney Institute for Astronomy; School of Physics; University of Sydney; NSW 2006; Australia
[7]Laboratoire d'Astrophysique de Grenoble; UMR 5571 Université Joseph Fourier/CNRS; BP 53; 38041 Grenoble Cedex 9; France

## 1. Introduction

Within the last 10 years, long-baseline optical interferometry (LBOI) has benefited significantly from increased sensitivity, spatial resolution, and spectral resolution, e.g., measuring the diameters and asymmetries of single stars (Peterson *et al.* 2008; Tatebe *et al.* 2007), imaging/fitting the orbits of multiple stars (Hummel *et al.* 2003), modeling Be star disks (Kanaan *et al.* 2008), and modeling AGN nuclei (Beckert *et al.* 2008). Similarly, polarimetry has also yielded excellent astrophysical results, e.g., characterizing the atmospheres and shells of red giants/supergiants (Beiging *et al.* 2006), modeling the envelopes of AGB stars (Gledhill 2005), studying the morphology of Be stars (Wisniewski *et al.* 2007), and monitoring the short- and long- term behavior of AGNs (Moran 2007). The next logical evolutionary step in instrumentation is to combine LBOI with polarimetry, which is called optical interferometric polarimetry (OIP). In other words, measurements of spatial coherence are performed simultaneously with measurements of coherence between orthogonal polarization states.

Based on the advancements listed above, we expect that OIP will provide new and exciting insights in the fields of stellar and extra-galactic astronomy when polarizing structures are spatially resolved (Section 2). A national LBOI facility consisting of a significant number of large telescopes will be built in the United States within the next 10-20 years. Space-based interferometers are being considered as well. The goal of this white paper is to provide the background summary for an OIP roadmap – including scientific justifications, technical challenges, and calibration issues – which in turn will be used to incorporate high-sensitivity OIP capabilities into next-generation instruments.

## 2. Scientific Justification

OIP is a powerful probe of circumstellar scattering environments that contain ionized gas or dust. Spatial asymmetries and magnetic confinement/alignment may play a part as well. Even stars that exhibit no polarization because of spatial symmetry when observed with classical polarimeters exhibit non-zero polarization when observed with OIP, which means more information is available to constrain atmospheric models. This section contains a partial list of OIP targets: young stellar objects, main sequence stars, red giant/supergiant stars, luminous blue variables, Be stars, Algol binary stars, AM Her binary stars, and Seyfert galaxies.

*2.1 Young Stellar Objects*

These stars have just begun their lives on the main sequence. Their circumstellar environments contain jets and disks; the latter are the birthplaces of planetary systems. Some disks have already been observed via LBOI (Monnier and Millan-Gabet 2002.).

The polarization signatures of disks depend on location and wavelength, which means that spatially and spectrally resolved OIP observations will provide clues to the composition, size, and number density distribution of dust grains. These parameters, in turn, act as inputs to planet formation models.

*2.2 Main Sequence Stars*

Monnier, Zhao, Pedretti, *et al.* (2007) resolved the surface of Altair, measuring temperature contours for the first time. The atmospheric temperatures range between 7000-8000K from equator to pole. The exact temperatures depend on gravity darkening and differential rotation models, which must be constrained with other types of measurements.

Normal spherical stars exhibit no intrinsic polarization when observed with a single telescope because circular symmetry cancels the net polarization. LBOI breaks this symmetry when the object is resolved, so an instrument outfitted with OIP can measure atmospheric polarization (Figure 1). As a result, atmospheric models can be constrained further, leading to a greater understanding of scattering mechanisms, line-formation mechanisms, and temperature/density contours and profiles for different atomic and molecular species.

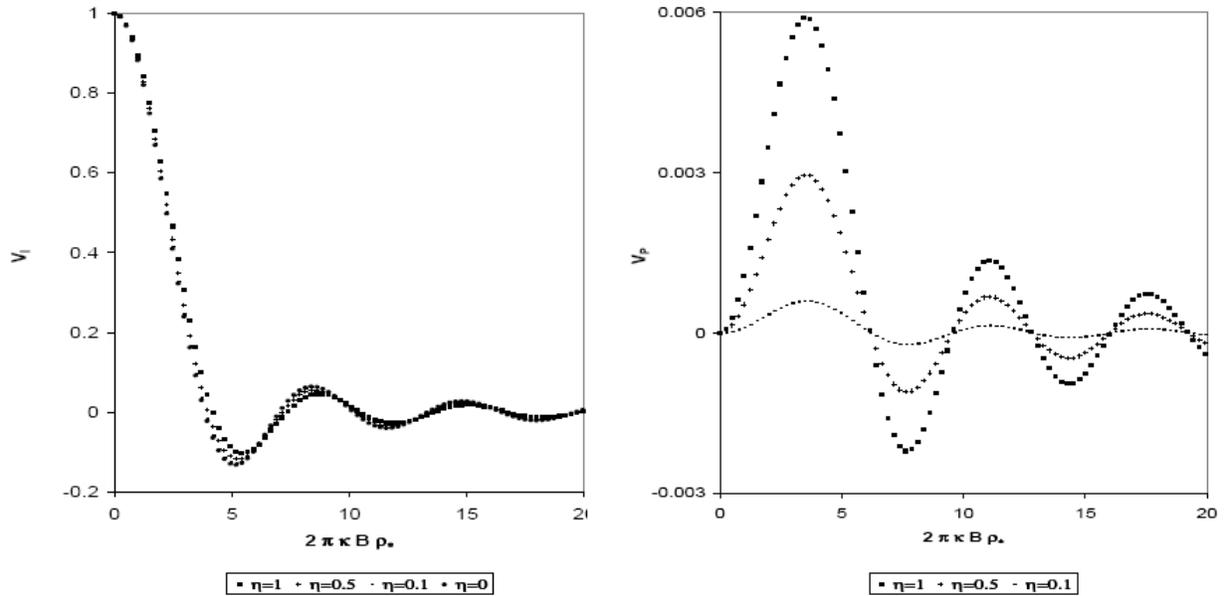

**Figure 1.** Stokes-I and linear-polarization visibility plots for an idealized Thompson-scattering star with a gray atmosphere. The three curves correspond to different fractions of a Thompson scattering atmosphere. Many different atmospheres can produce the curves on the left, but OIP provides additional information to narrow the number of possibilities. From Elias (2004a).

Some stellar atmospheres are not perfectly symmetric. The most extreme cases are Ap and Bp stars with kilogauss magnetic fields and peculiar chemical abundances (Shulyak *et al.* 2008). They should show significant asymmetric polarizing structures when observed with OIP, dependent on how the magnetic fields align/confine the atmospheric gas. Donati, Morin, Petit, *et al.* (2008) studied polarized line profiles to tomographically reconstruct the magnetic field structures of M dwarfs, which should also be interesting when observed with OIP.

*2.3 Red Giant/Supergiant Stars*

Ireland, Tuthill, Davis, and Tango (2005) performed OIP observations of the Mira red giant stars R Car and RR Sco using the Sydney University Stellar Interferometer (SUSI). The OIP was limited to modeling visibilities in perpendicular polarization states (no Stokes parameters). The polarized flux of the light scattered by dust is much fainter than the photospheric flux. The inner radius of dust formation is less than ≈ three stellar radii, much smaller than expected for "dirty silicate" grains. A geometrically thin shell fit the data better than an outflow.

Red supergiant stars represent the final stage of massive star evolution. The photospheres of the objects are highly convective, contain many molecular species, and may exhibit hot/cool spots. Red supergiants have been resolved with classical LBOI. Their diameters depend on wavelength and can vary greatly between line and continuum (Dyck and Nordgren 2002). Temporal variability has been found via classical polarimetry (Hayes 1984; Holenstein 1991), suggesting the existence of slowly evolving polarizing structures. OIP measurements will spatially resolve these polarizing structures.

*2.4 Luminous Blue Variable Stars*

Luminous blue variables (LBVs) are extremely massive early-type supergiants that exhibit occasional episodes of substantial mass loss superimposed on slow brightness changes. The most well known and well studied of these stars is η Car, which has been observed with classical polarimetry (King 2002), imaging polarimetry at a spatial resolution of ~ 0.2" (Walsh and Ageorges 2000), and LBOI at spectral resolutions of 1500 and 12000 (Weigelt 2007). The results were consistent with a rapidly rotating hot star and enhanced mass loss at the poles. OIP at high spatial resolution will provide new information about the distribution and constituents of the dust and gas as well as their dynamics.

*2.5 Be Stars*

Be stars are B stars with extensive ionized disks. They exhibit strong hydrogen emission lines and linear polarization perpendicular to the disk at the ~ 1% level. These objects were observed at Hα with an LBOI, verifying the oblate nature of the disks (Quirrenbach *et al.* 1997). OIP simulations have shown that complex Stokes visibilities depend strongly on model parameters (Figure 2), such as inclination, density profile, opening angle, etc. Despite decades of study, the physical processes that form and maintain Be star disks are not well understood, which represents the major unsolved puzzle of this field. OIP images will lead to information about the hydrodynamics and temporal evolution of these disks so that theoretical models can be constructed and tested.

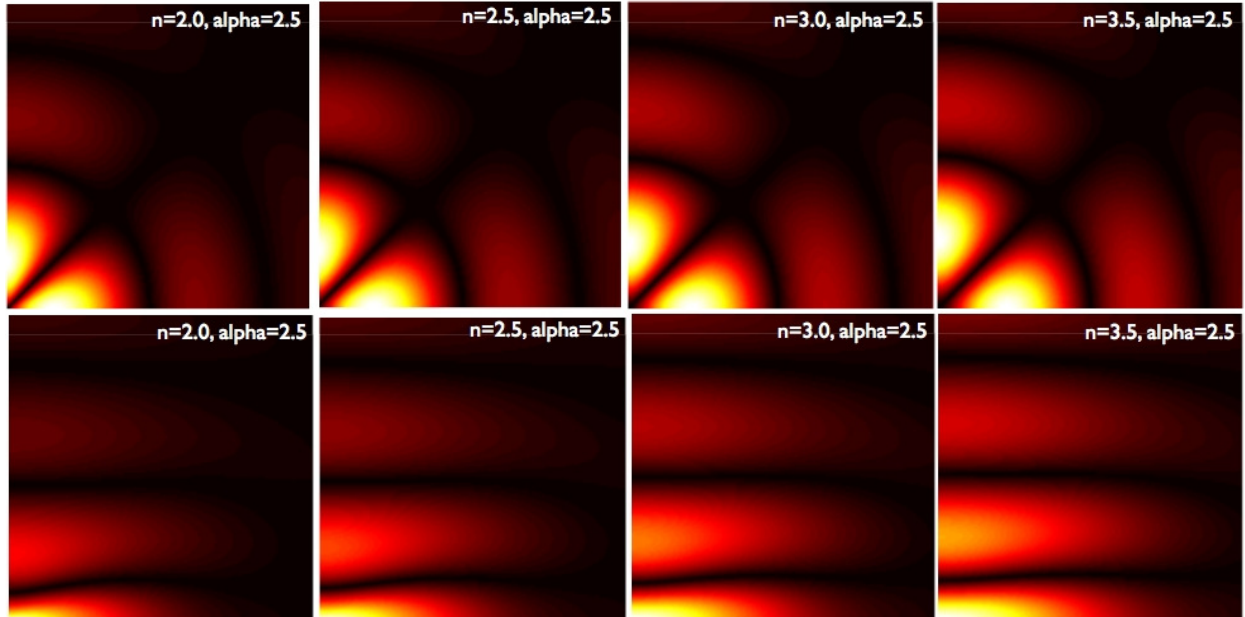

**Figure 2.** Stokes-Q visibility magnitudes for a Be star (the upper right quadrant of the Fourier plane). The bright regions correspond to values on the order of a few tenths of a percent. The top row represents a face-on (i=0°) system, while the bottom row represents an i=60° system. The columns (l-r) represent increasing number density of electrons. The opening angle of the bow-tie disk is alpha=2.5°. These countour plots are significantly different from circularly symmetric Stokes-I visibility magnitudes. Also, the i=0° and i=60° countour plots are significantly different, making modelling and imaging easier compared to Stokes-I visibility magnitudes. From Mackay, Elias, Jones, and Sigut (2008).

*2.6 Algol Binary Stars*

Algols are mass-transferring binary stars. The stellar types and luminosity classes vary over a wide range, depending on the evolutionary state. The class prototype Algol exhibits phase-locked and sporadic light curves (Kim 1989) and phase-locked polarization curves produced by Thompson scattering in the mass-transfer stream (Wilson and Liou 1993; Elias, Koch, and Pfeiffer 2008).

Another well-studied Algol-type system is β Lyr. It is also the most complicated and misunderstood object of the class because it has the largest amount of circumstellar material. β Lyr has been observed polarimetrically by a number of workers (Figure 3). Unlike other Algol binaries, it exhibits a large spike in the position angle produced by the eclipse of the stream/disk impact region. It has also been observed via classical LBOI at limited spatial and spectral resolution (Schmitt *et al.* 2006), but even so the relative motion between the continuum and Hα emission was clearly observed. Recently, Zhao, Gies, Monnier, *et al.* (2008) resolved β Lyr at six different epochs using LBOI, clearly resolving the mass-losing star and the accretion disk surrounding the mass-gaining star.

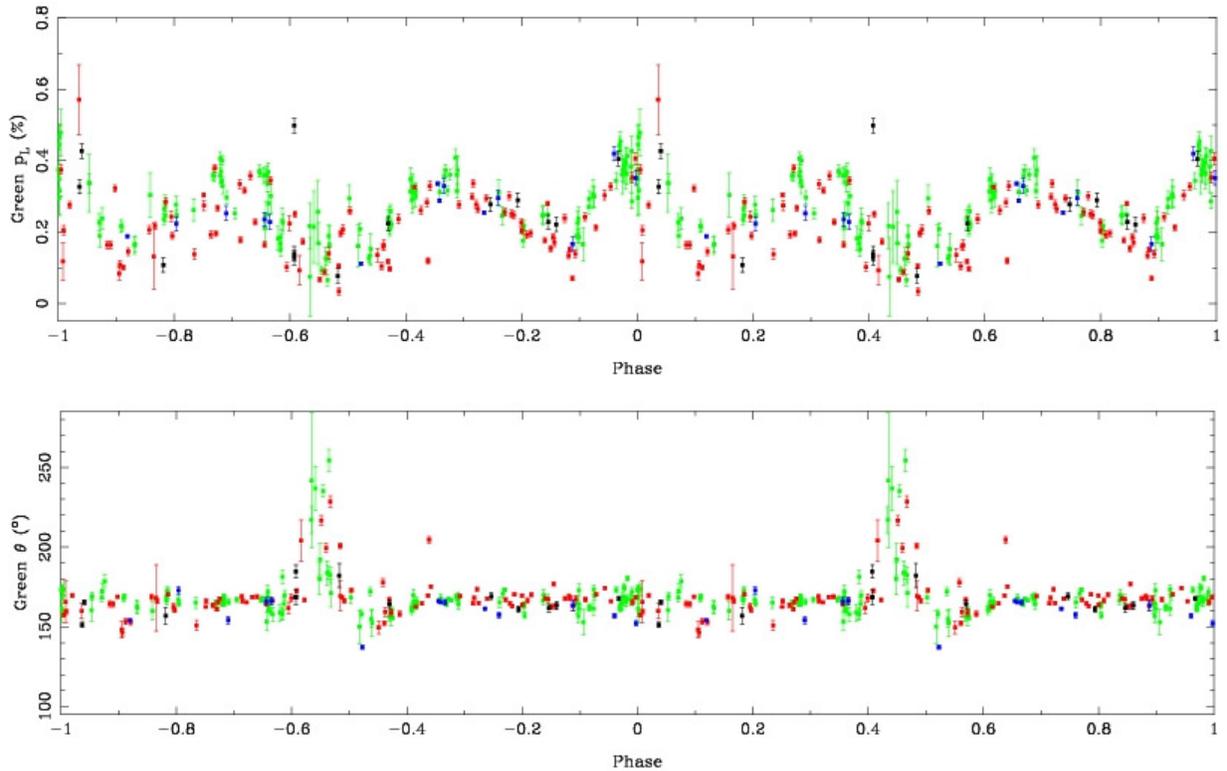

**Figure 3.** Classical linear polarization magnitudes (top) and position angles (bottom) of β Lyr in the Johnson V filter. Note the prominent peak of the magnitude near phase=0 (disk in front of primary star) and the peak of the position angle near phase=0.45 (streak/disk impact eclipsed). The green points come from Appenzeller and Hiltner (1967). The red points are data from Elias, Koch, and Holenstein (1996). The black and blue points were published by Hoffman, Nordsieck, and Fox (1998).

Many questions remain about the nature of β Lyr. What is the trajectory of the mass-transfer stream? What is the nature of the optically thick component of the accretion disk? What does the steam-disk impact region look like? Are there scattering mechanisms other than Thompson scattering? Is there a systemic wind? How does β Lyr, among the most active of Algols, fit into the evolutionary sequence of close binary stars? High spatial and spectral resolution OIP will answer these questions.

*2.7 AM Her Binary Stars*

These evolved close-binary stars consist of a white dwarf and a late main-sequence dwarf that has reached its Roche lobe. The red dwarf loses mass directly to the poles of the white dwarf because of its mega-gauss magnetic field, as opposed to other cataclysmic variables (CVs) where the mass accretes onto a disk. The orbital periods are short, on the order few hours.

AM Her objects exhibit phase-locked linear and circular polarization, often as large as a few tens of percent peak-to-peak (Figure 4). Polarization observations have been used to model the magnetic field structure near the poles. OIP will extend these results by mapping the mass flowing along the magnetic field lines.

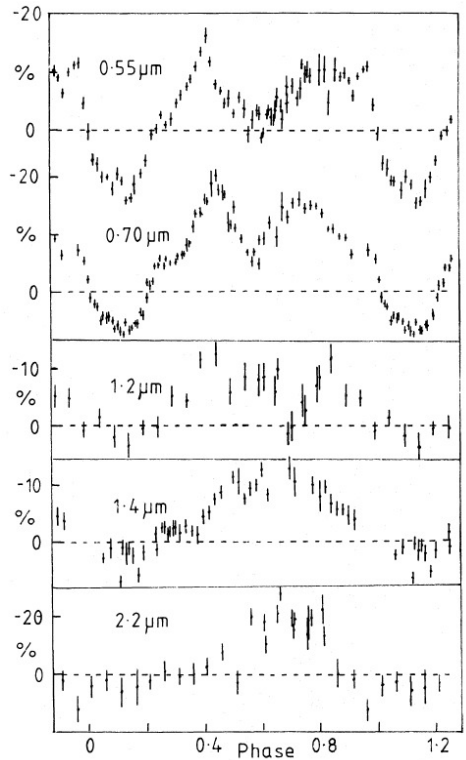

**Figure 4.** Circular polarization measures of AM Her from the visible to the near infrared. Most of the signal arises from the orientation of the cyclotron radiation from the accretion streams with respect to the observer. The data come from Bailey, Hough, Gilmozzi, and Axon (1984).

*2.8 Seyfert Galaxies*

Present models suggest that Type I and Type II Seyferts are intrinsically the same, a black hole surrounded by a small and rapidly orbiting disk plus an extended region. Type II Seyferts produce polarization at arcsecond scales, resolvable by HST. Type I Seyferts, however, have reflection and scattering regions closer to the nucleus.

Two of these galaxies have already been observed using LBOI at 10 µm (Poncelet, Perrin, and Sol 2006). OIP will provide enough information to verify the simplified model, determine the exact scattering mechanisms, and show if there are any intrinsic differences between Type I and Type II objects.

# 3. Practical Issues

The sample of objects discussed in Section 2 clearly show that OIP can provide significant new insights for many types of objects. What can the astronomical community do to insure that future ground-/space- based LBOI facilities will include OIP capabilities? In this section, we list practical issues associated with this question.

*3.1 Optics*

There are two main optics issues related to OIP. The most important is the calibration of instrumental polarization. The other, design of beam combiners, will determine how OIP is incorporated into LBOI systems. We believe that initial OIP testbed systems should employ low to moderate spectral resolution, to maximize S/N.

3.1.1 Instrumental Polarization

Since existing interferometers employ mirrored feed systems, the most important issue is instrumental polarization. It must be removed, otherwise the OIP data will be misinterpreted. Instrumental polarization is large at optical wavelengths, on the order of 1-10%, much larger than a single prime-focus or Cassegrain telescope designed for classical polarimetry. It should be somewhat smaller at near-infrared wavelengths. The effects of instrumental polarization of classical LBOI and OIP measurements have been discussed (Elias 2001; Elias 2004b) and modeled as a function of pointing (Figure 5), but more work is required. An instrumental polarization compensation device was tested successfully at GI2T (Rousselet-Perraut *et al.* 2006), which may be useful for future OIP designs.

Additional instrumental polarization questions for mirrored feed systems include:
- How do source and instrumental polarization (for both matched and mismatched arms) affect the complex visibilities measured by classical interferometry?
- Is there an optimum telescope and/or feed system configuration for OIP? MROI makes a good attempt at maintaining polarization fidelity (Buscher *et al.* 2008).
- What are the calibration models for each type of observable? Can model fits include pointing direction and field rotation?

- A significant number of classical polarization calibrators can be found in the literature. If they are resolved with an interferometer (Section 2.2), OIP calibration must then include modeling versus baseline length. Can this complication be mitigated?
- Do OIP instruments depolarize significantly? If not, in principle it would be possible to calibrate Stokes visibilities based on separate classical calibrations of each arm. The polarization structures of calibrator stars then would not be resolved (see previous bullet), which means that OIP calibration will be much simpler.
- If OIP instruments do depolarize significantly, can this effect be overcome?
- Will atmospheric turbulence significantly increase systematic errors in some types of polarized visibility observables? Will a separate scalar calibration, as in classical interferometry, mitigate this problem?
- Can OIP calibrators be observed sporadically (to maintain a previously determined instrumental polarization calibration versus pointing) or must they be observed along with the program stars?

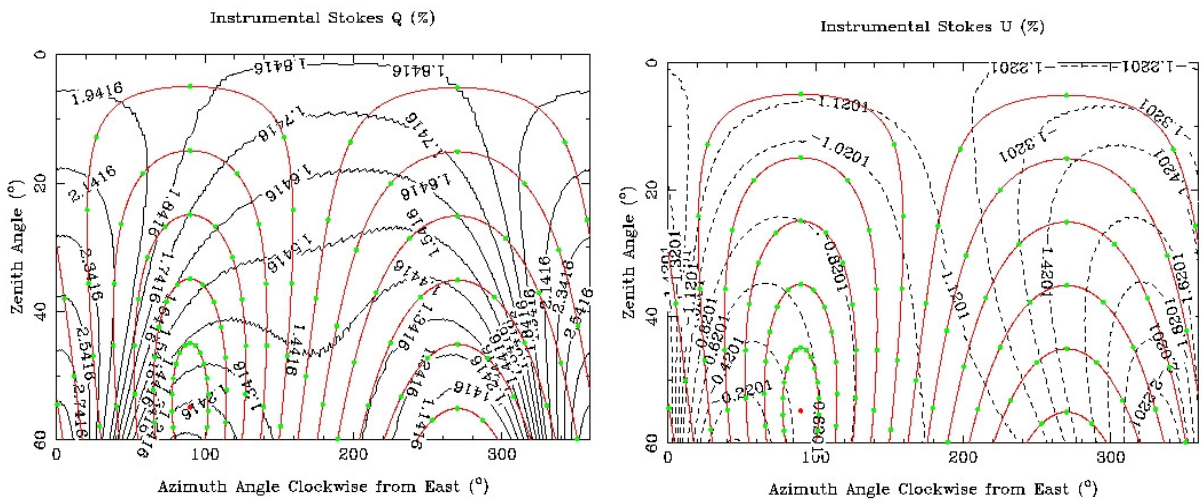

**Figure 5.** Non-interferometer instrumental polarization (normalized Stokes Q and U) of NPOI as a function of pointing (Elias 2002). The black lines are polarization contours (%). The red lines represent declination contours in 10° decrements; 90° is red dot on the lower left of each plot, and the contours on the right are south of 40°. The simulations assume green light and silver mirrors, excluding the beam combiner.

It is possible that future optical interferometers will employ fiber instead of mirrors for their feed systems. The questions listed above are also relevant for fibers, but there are other issues that should be addressed as well:
- How do various types of fibers affect polarization? Do any fibers truly preserve polarization (Le Bouquin et al. 2005)?

- Are single-mode fibers required?
- Since OIP requires relatively high S/N, can injection and propagation losses in fibers be minimized?

3.1.2 Beam Combiners

Instrumental polarization contains linear and circular components that interact with each other. To insure properly calibrated OIP observations, the beam combiner should be able to simultaneously measure both linear and circular polarization (i.e., an elliptical polarimeter). There are also compelling science arguments for elliptical polarimetry: evolved stars with dust shells and AM Her objects (Section 2) are significant sources of both linear and circular polarization.

Existing beam combiners can be modified for OIP by installing a polarimeter at the output (Elias 2004b). The present NPOI beam combiner has an unused output that could be employed for this purpose. Calibration may be difficult, but the experience will be useful for future instrument design.

Future beam combiners might integrate polarizing optics along with the rest. There are many other possible designs, depending on the application. Vega/CHARA will measure dispersed visibilities for each polarization state as well as the phase difference between them (limited OIP; Mourard *et al.* 2008). A more complicated instrument could employ circular polarizers (left, L; right R) and non-polarizing beam splitters to correlate LL*, LR*, RL*, and RR*, which in turn can be combined to form the Stokes visibilities. This technique is employed at radio wavelengths with circular feeds. To increase the S/N as well as obtain amplitudes and phases of the Stokes visibilities, the beam-combiner and fringe-tracker design should promote coherent averaging (Jorgensen *et al.* 2008).

Fully integrated beam combiners on chips (Kern *et al.* 2008) represent a possible new direction for beam combiners. They are small and can be constructed for any beam combination scheme. Can the channels within the chips be modified to separate orthogonal polarization states? If so, the "radio-like" beam combiner mentioned in the previous paragraph could be constructed with relatively little effort.

*3.2 Data Reduction*

Data reduction issues include the OIP observables, interstellar polarization, software, and data formats.

3.2.1 Observables

The observables for classical polarimetry are the normalized Stokes parameters $q_0 = Q_0/I_0$, $u_0 = U_0/I_0$, and $v_0 = V_0/I_0$. At radio wavelengths, the observables are the correlated Stokes fluxes $I_{12}$, $Q_{12}$, $U_{12}$, and $V_{12}$. In near-infrared and optical wavelengths, we must employ normalized OIP quantities, just as for classical polarimetry, because the photometry is not as stable. The exact form of the OIP observables depends on the capabilities of the beam combiner and the science goals.

It is possible to model sources using only the visibilities of the orthogonal polarization states and their relative phase (limited OIP; Mourard *et al.* 2008). To examples of observables that are related to Stokes parameters are (Elias 2004b): 1) the complex normalized Stokes visibilities $V_I = I_{12}/I_0$, $V_Q = Q_{12}/I_0$, $V_U = U_{12}/I_0$, and $V_V = V_{12}/I_0$; and 2) normalized Stokes parameters $q_{12} = Q_{12}/I_{12}$, $u_{12} = U_{12}/I_{12}$, and $v_{12} = V_{12}/I_{12}$. Both sets approach the normalized uncorrelated Stokes parameters as the baseline length approaches zero. The second set has the advantage of being less susceptible to atmospheric phase errors (no scalar calibration is required), but they also diverge when the object is highly resolved. NB: OIP measures both uncorrelated (identical to the classical normalized Stokes parameters) and correlated observables. The former contain useful "zero-spacing" information and should definitely be employed in data reduction.

3.2.2 Interstellar Polarization

Like instrumental polarization, interstellar polarization must be removed from the data before they can be interpreted. It is caused by forward scattering off interstellar dust grains. The dust grains are oblong or needle shaped, and they are aligned perpendicular to the galactic magnetic field. The induced interstellar polarization is thus parallel to the galactic magnetic field (Serkowski, Mathewson, and Ford 1975).

Many polarimetrists keep lists of null and non-null polarization standards. These stars tend to be uninteresting (main sequence, single, free of circumstellar material, etc.), just like interferometric standard stars. Therefore, these interstellar polarization lists can be used for OIP. Ultimately, we suggest creating a master list and making it available on the OIP homepage (Section 3.4). If depolarization is not an issue, then these lists may be used "as is," otherwise the baseline-length dependence must be taken into account (Section 3.1.1).

3.2.3 Software and Data Formats

At the present time, there is no freely available software capable of reducing OIP data. Some algorithms in existing software, such as preliminary scalar calibrations (e.g., closure phases), may be useful for OIP, with or without modification. OIP-specific calibration will require either augmenting existing software or writing completely new software.

Once the OIP data are calibrated, they must be interpreted via imaging or modeling. Some radio interferometry techniques, such as CLEAN or self calibration, could possibly be employed for OIP imaging, depending on the observables used. OIP modeling software must be created specifically for each type of object.

The present OIFITS file standard and the associated C library (Young *et al.* 2008) cannot handle OIP data. Radio interferometer arrays employ standard file formats. Can they be used as a starting point? Unfortunately, each OIP instrument may employ different observables and it may not be possible to convert all of them to a common set. This issue will require careful consideration.

*3.3 Improving Results from Existing and Future non-OIP Instruments*

Polarization mismatches between the arms of a feed system reduce the observed scalar (non-OIP) visibilities by up to a few tenths of a percent (Elias 2001). Observations of standard stars are designed to remove this reduction, but if the mismatch changes rapidly between the program and calibrator star observations, the calibration will be imperfect. This systematic error is especially problematic when measuring the low visibilities of resolved objects. Existing and future interferometers, such as CHARA, NPOI, and MROI, can be used as testbeds to study these effects.

Polarization also affects interferometers designed to measure exosolar planets. The same polarization mismatches mentioned above allow additional stellar leakage to pass through nulling interferometers (Elias 2003; Elias, Draper, and Noecker 2004). For narrow-angle astrometry instruments, polarization mismatches introduce small shifts in the differential phases between the fringes of the stars, which in turn introduce systematic errors to the differential-delay corrections (Geisler *et al.* 2008). PRIMA/VLTI could be used as a testbed to study these effects.

Space-based interferometers (non-nulling and nulling) will have very high quality optics. Also, they are not hindered by the Earth's atmosphere. It is logical, therefore, to think of add-on instrumentation for ancillary science projects, to get the highest possible return on investment from the mission. OIP instruments are good choices. The scientific results will be significant (Section 2). The instrumental polarization should be relatively small and nearly constant. Depolarization will be minimal. Plus, there will be no systematic scalar calibration errors due to atmospheric turbulence.

*3.4 Communication*

A dedicated OIP web page is under construction at http://www.lsw.uni-heidelberg.de/users/nelias/OIP.html. It will contain OIP news, an OIP bibliography, public OIP proposal documents, and links to public OIP data.

Communication among interested parties will be conducted via private e-mail or the OLBIN majordomo mailing list, at least for now. OIP breakout sessions should be scheduled at major conferences. When enough interest is generated, perhaps a temporary working group under IAU Commission 54 can be created as well.